\documentclass[twocolumn,twoside,slac_two,floatfix]{revtex4}
\usepackage{graphicx}
\usepackage{fancyhdr}
\usepackage{multirow}
\usepackage{url} 
\usepackage{hyperref}
\begin{document}

\title{Temporal properties of bright BGO GRBs detected by Fermi}

%
\author{E. Bissaldi$^{\rm \,1,2,3}$, E. Peretti$^{\rm 2}$, F. Longo$^{\rm 2,3}$}
\affiliation{$^{\rm 1}$INFN -- Sez.~di Bari, Via E. Orabona 4, 70125 Bari, Italy \\ $^{\rm 2}$Physics Dep., University of Trieste, Via Valerio 2, 34127 Trieste, Italy \\ $^{\rm 3}$INFN -- Sez.~di Trieste, Via Valerio 2, 34127 Trieste, Italy}
%
%
\author{On behalf of the Fermi LAT Collaboration}
\affiliation{  }
\begin{abstract}
We present results of an analysis of a sample of bright Gamma-Ray 
Bursts (GRBs) detected by Fermi-GBM up to more than 1 MeV, 
which were collected during six years of Fermi operations. 
In particular, we focus on the GRB durations over several 
energy bands of the prompt emission of a subsample of bright 
GRBs detected up to 10 MeV by GBM and, when possible, 
up to 1 GeV by Fermi-LAT, 
thus expanding the Duration--Energy relationship in GRB 
light curves to high energies for the first time.
We find that the relationship for these energetic GRBs 
is flatter than reported for other samples, 
suggesting that the high-- and low--energy emission 
mechanisms are closely related.
\end{abstract}
%
\maketitle
\thispagestyle{fancy}
%
%
\section{Fermi GBM and LAT instruments}
\begin{figure}[b!]
\centering
\includegraphics[width=65mm,bb=0 0 156 187,clip]{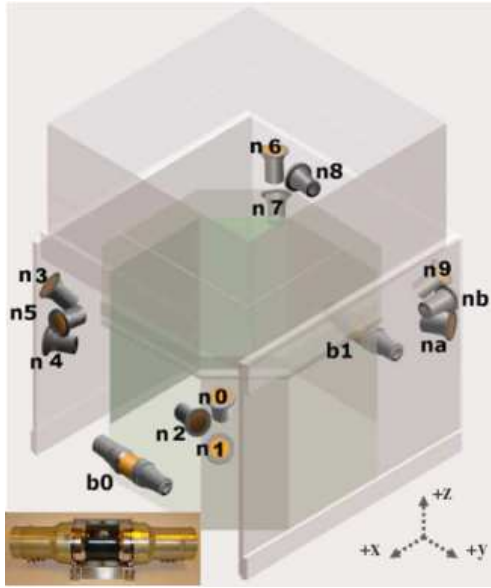}
\caption{Schematic view of the Fermi GBM detectors. The insert
in the bottom left corner shows one of the two BGO detectors.} 
\label{Fig1}
\end{figure}

The Fermi satellite has been observing the gamma--ray sky
since its launch in June 2008. 
It carries on--board two instruments:
the Gamma-ray Burst Monitor (GBM) 
and the Large Area Telescope (LAT). 
GBM consists of 12 Sodium Iodide (NaI, 8--900 keV)
and 2 Bismuth Germanate (BGO, 200 keV--40 MeV) 
scintillation detectors \cite{MEE09}.
Figure \ref{Fig1} shows the placement and 
orientation of the detectors on the spacecraft,
which allow GBM to have a Field-of-View (FoV)
as large as the full unocculted sky.
GBM detects $\sim$250 GRBs per year \cite{VON14}.

The LAT instruments include a Tracker-Converter, 
a Calorimeter and an Anti--Coincidence Detector \cite{ATW09}.
The LAT standard analysis covers an energy range 
of 100 MeV--300 GeV. Thanks to the LAT Low Energy 
(LLE) technique \cite{PEL10}, this coverage is extended 
down to 10 MeV. 35 GRBs were observed by LAT during the
first 3 years of operation \cite{ACK13}, but many more are expected
to be found thanks to a new analysis algorithm \cite{VIA14}
and to the newly implemented LAT event 
reconstruction {\it Pass 8} \cite{ATW13}.
\begin{figure*}[t!]
\centering
\begin{tabular}{cc}
\includegraphics[width=65mm,bb=0 0 595 504,clip]{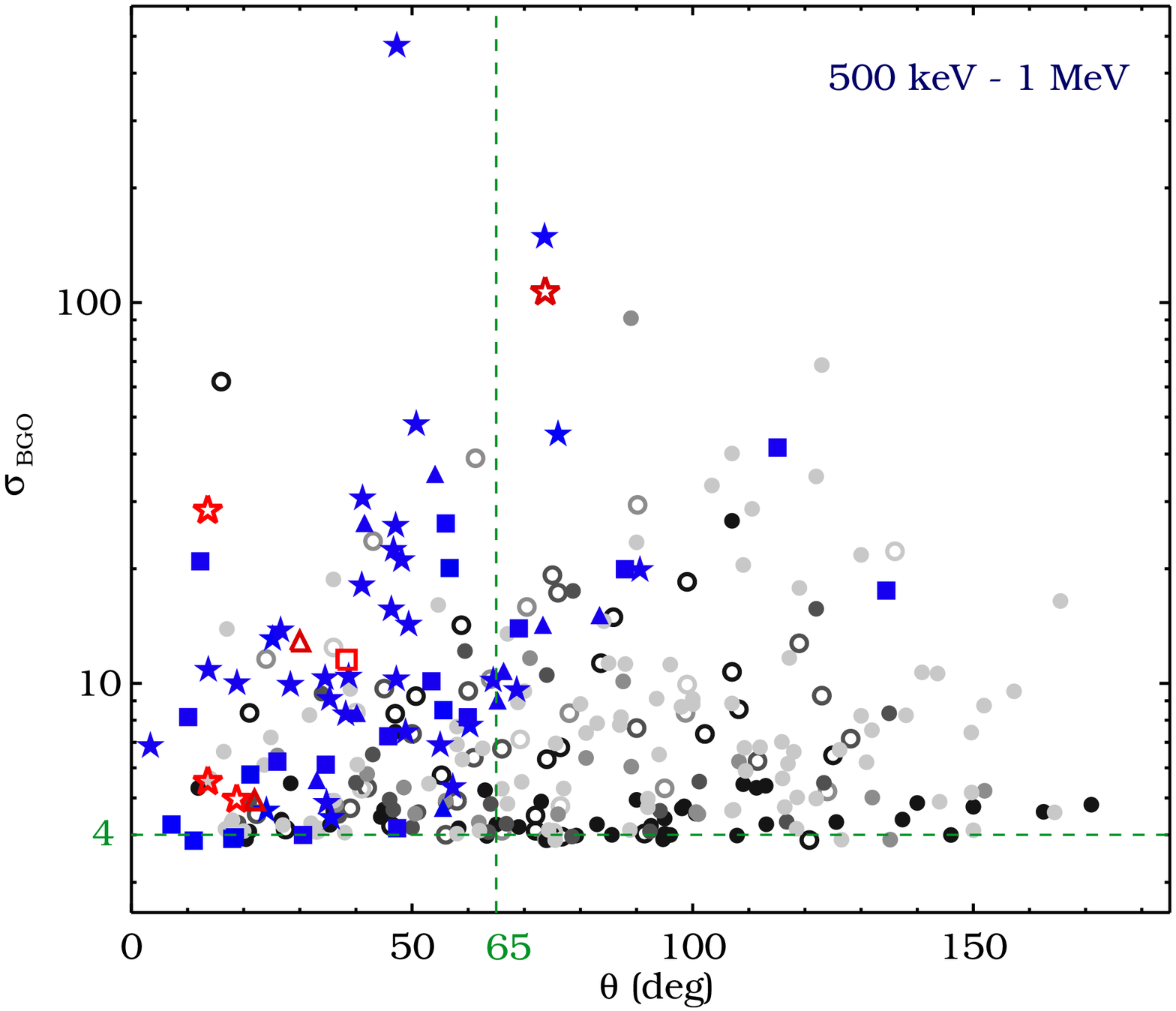} & \includegraphics[width=65mm,bb=0 0 595 504,clip]{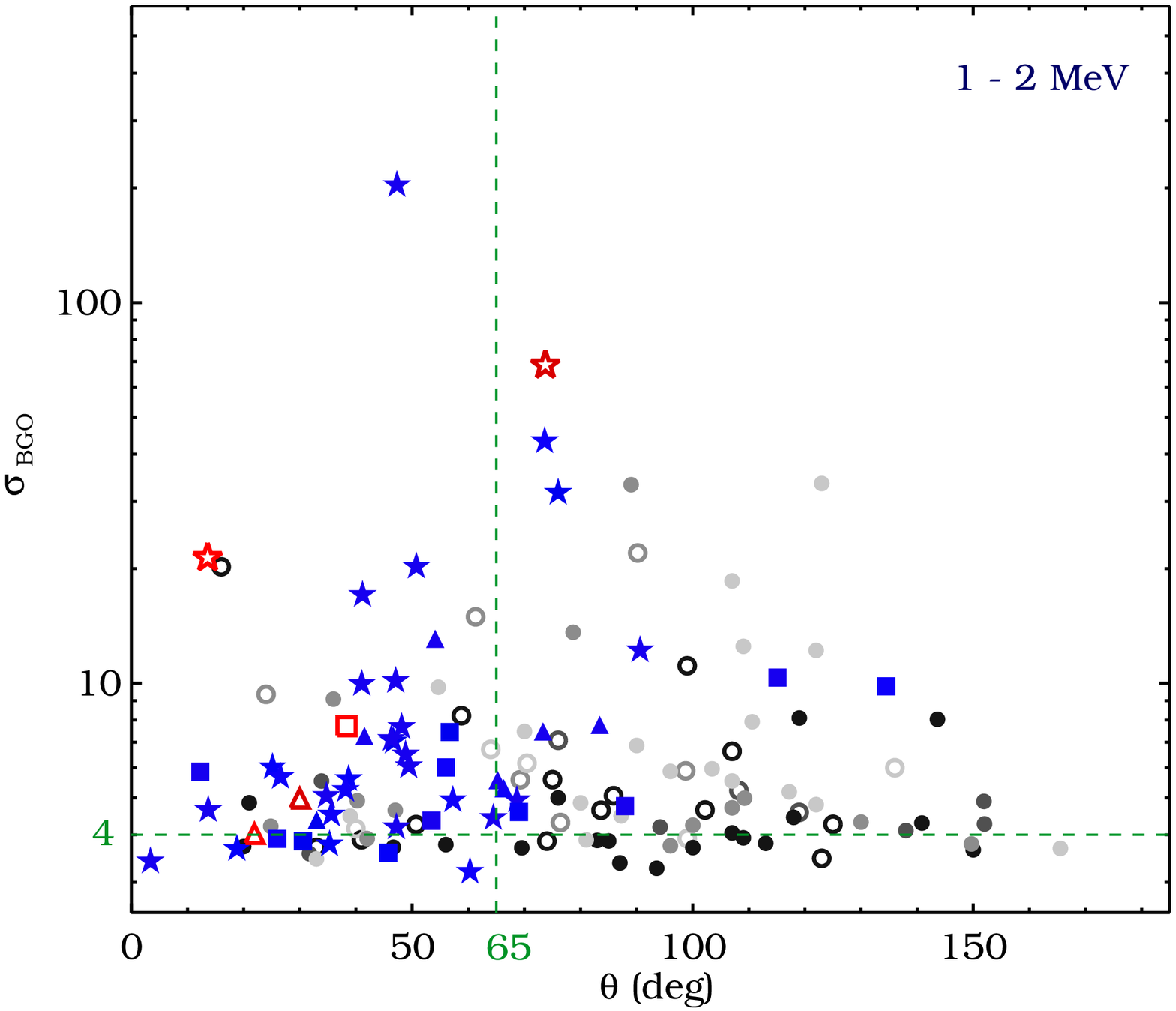} \\
\includegraphics[width=65mm,bb=0 0 595 504,clip]{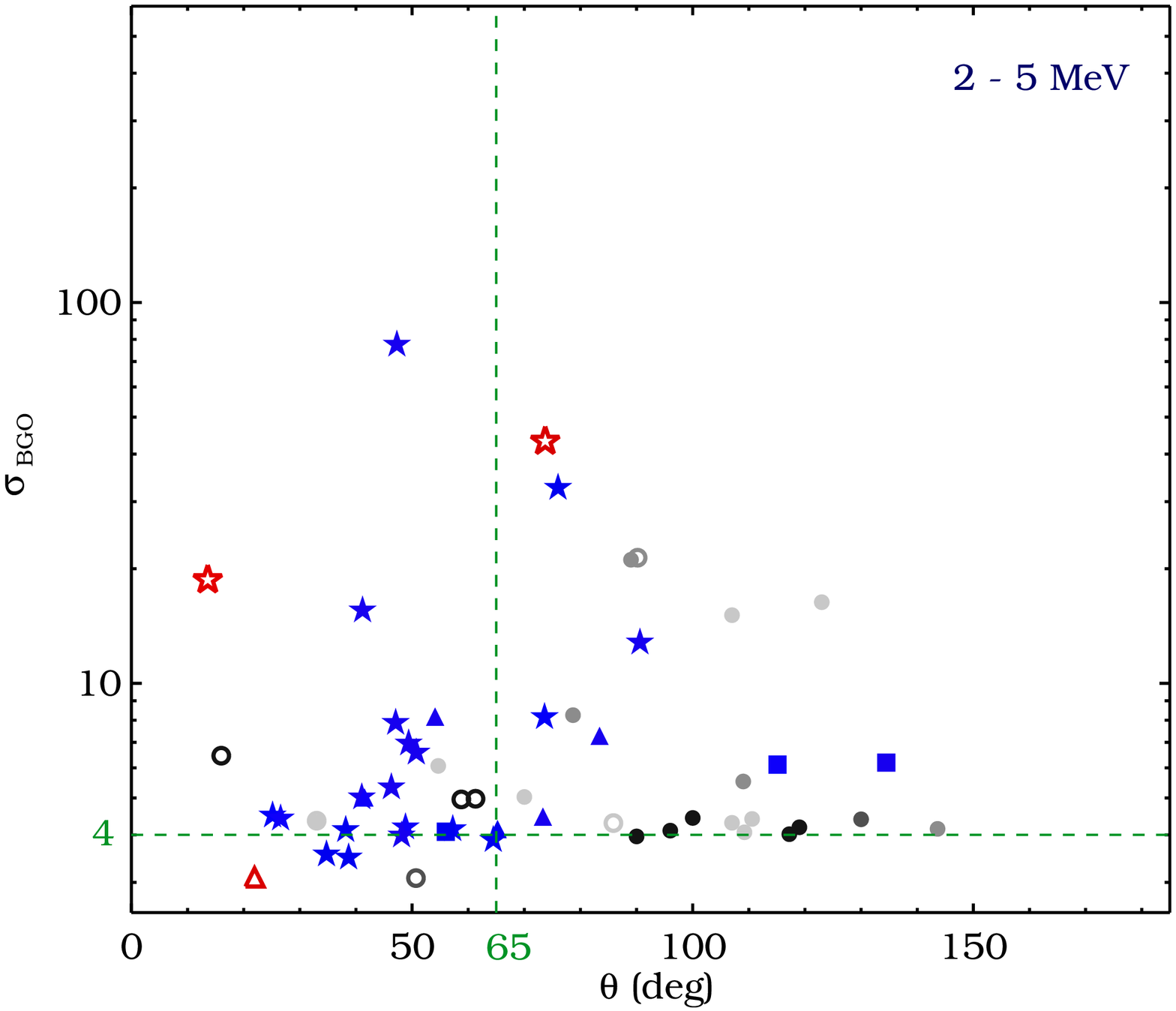} & \includegraphics[width=65mm,bb=0 0 595 504,clip]{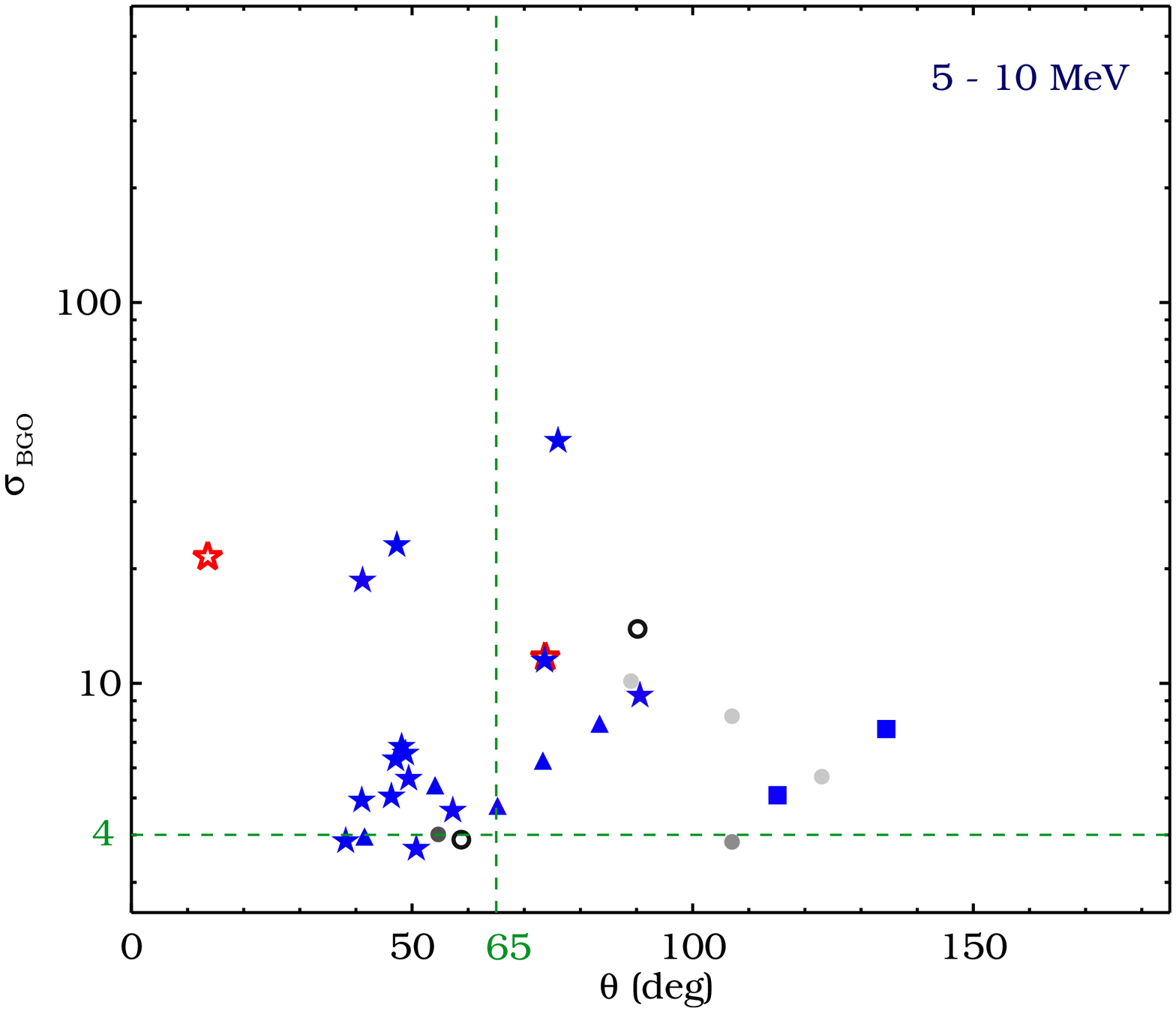}
\end{tabular}
\caption{BGO detection significance versus LAT boresight angle
$\theta$, calculated in four BGO energy ranges (see top right
corner of each plot).} 
\label{Fig2}
\begin{tabular}{cc}
\includegraphics[width=65mm,bb=0 0 595 504,clip]{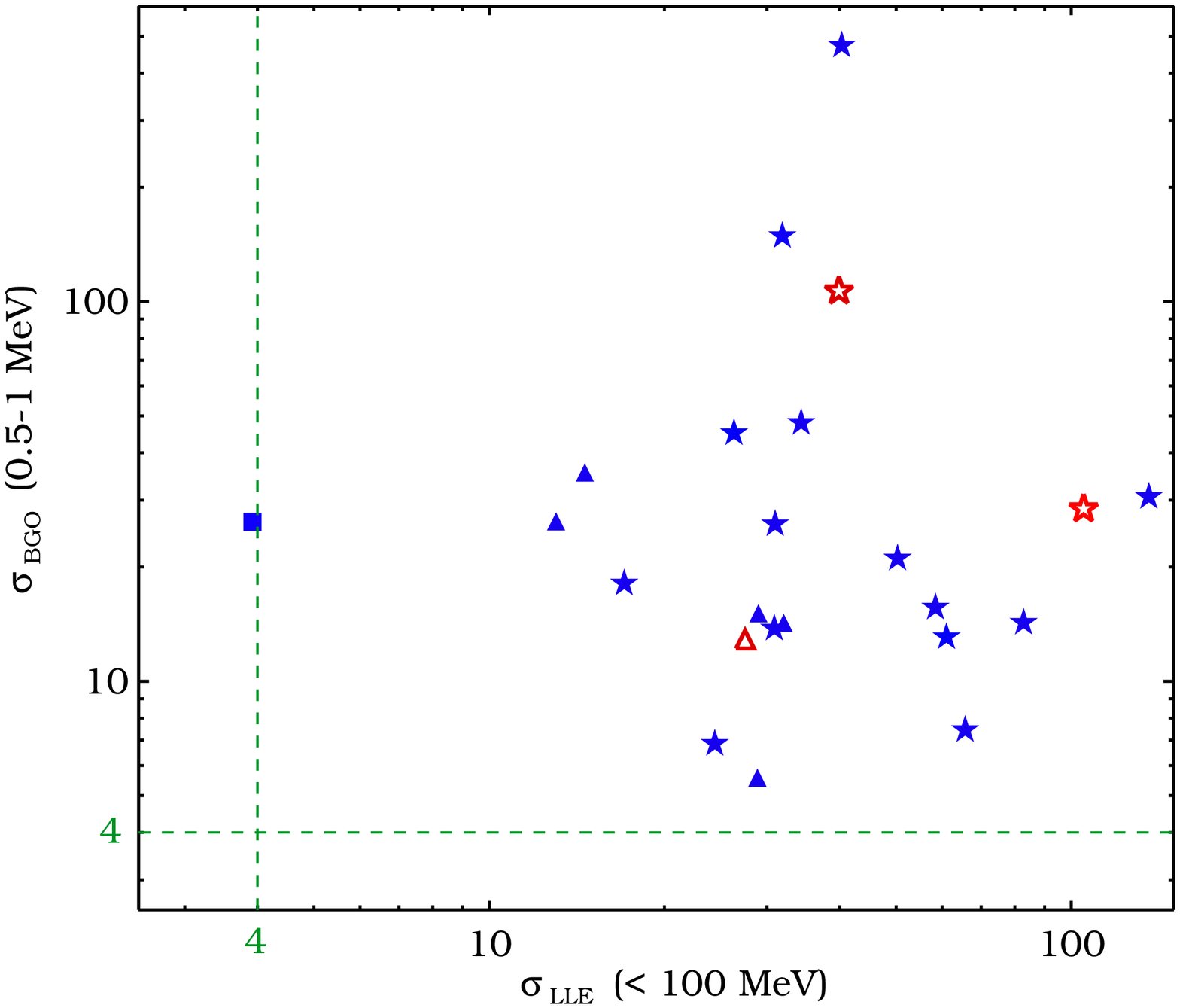} 
\includegraphics[width=65mm,bb=0 0 595 504,clip]{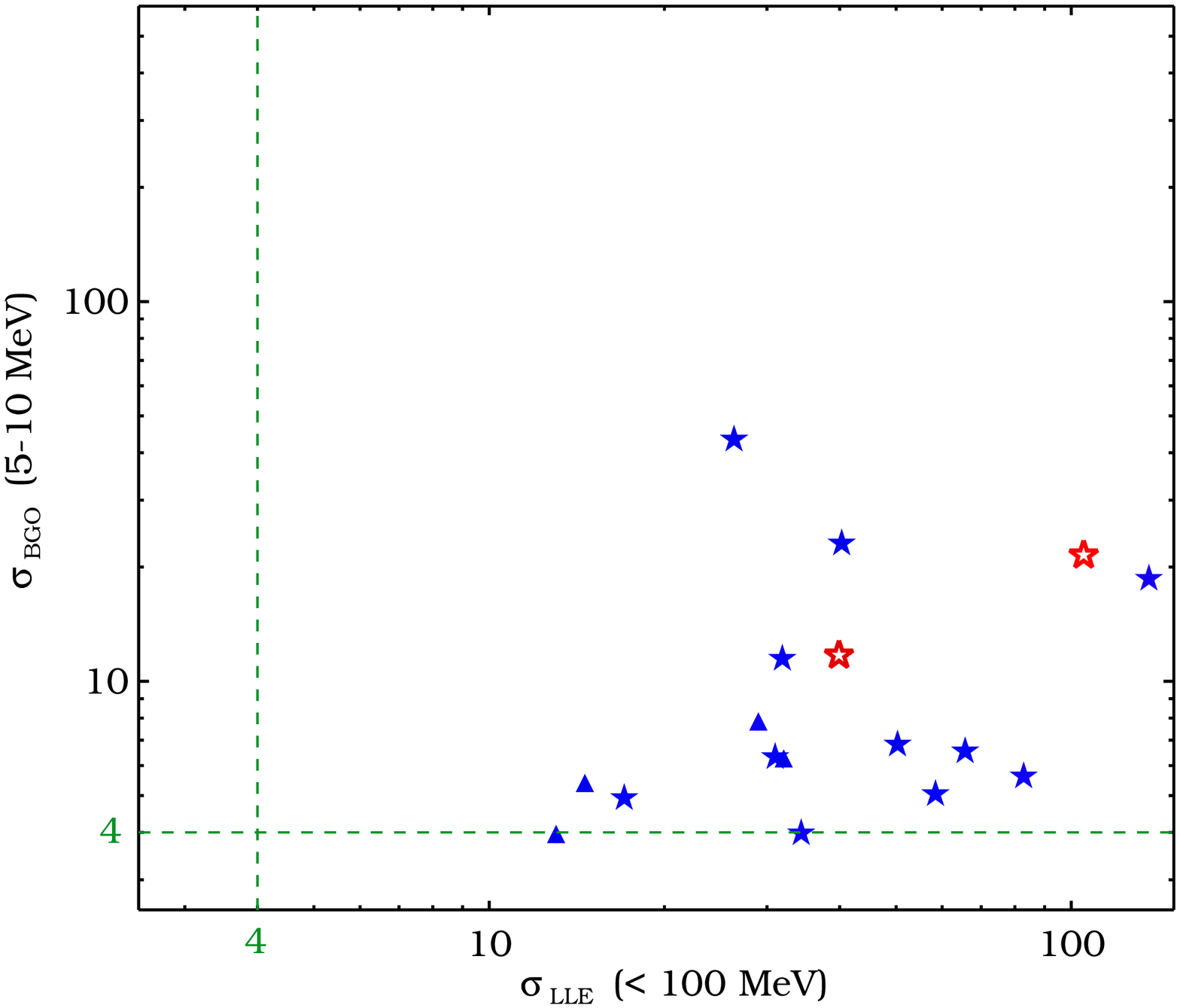} 
\end{tabular}
\caption{BGO versus LLE detection significances for two 
different BGO energy ranges: 0.5--1 MeV ({\it left panel} and 5--10 MeV
({\it right panel}).} 
\label{Fig3}
\end{figure*}

\section{Sample selection criteria}\label{sec2}
We follow the same approach as in \cite{RIC96}
for the BATSE bright GRBs and in
\cite{BIS11} for the GBM BGO bright GRBs collected
over the first year of operation. Here we 
extend the previous analysis to six years of GBM 
data (from August 2008 to July 2014).

The first coarser selection is based on the analysis of the 
GBM telemetry packets. Bursts showing an increase of 
more than 80 counts/s over background in at least one 
BGO detector over the full BGO energy range are selected. 
The second finer selection is based on the analysis of 
the count rate excess above background measured 
by the BGO detector(s) in the 500 keV--1 MeV 
range during the main burst emission episode.
We analyse the GBM TTE files (see \cite{MEE09} for
GBM data type description) 
over four different
timescales (64, 128, 256, and 512 ms).
Bursts with a 4$\sigma$ detection are selected. 

The final sample of {\it bright BGO GRBs} 
includes 311 bursts, of which 68 are short and 243 are 
long ones. We repeat the same procedure on three other
BGO energy ranges, namely  1--2 MeV, 2--5 MeV, and 5--10 MeV,
and check for the detection significance.

\begin{figure*}[t!]
\centering
\begin{tabular}{cc}
\includegraphics[width=70mm,bb=0 0 590 633,clip]{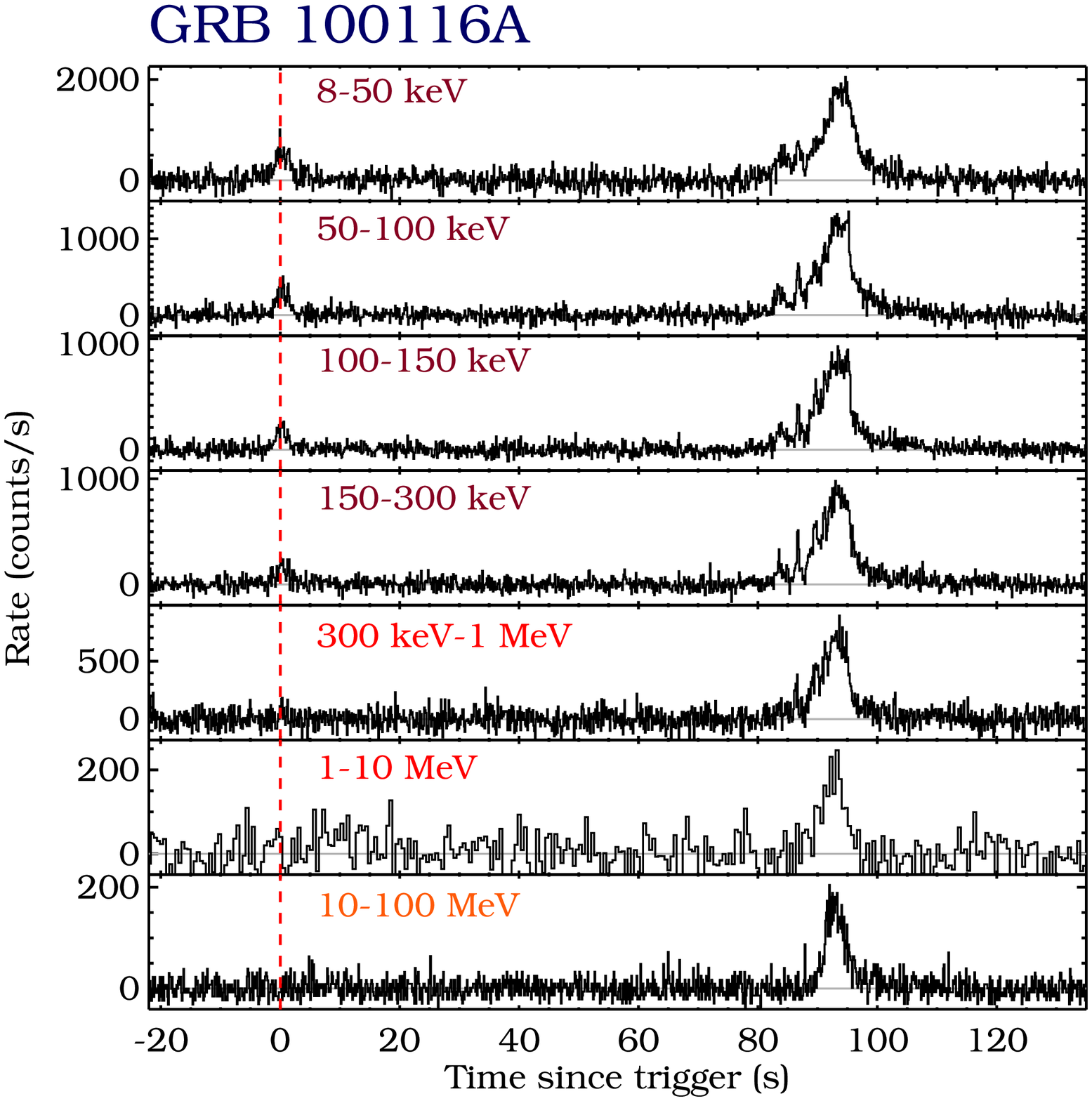} &
\includegraphics[width=70mm,bb=0 0 590 633,clip]{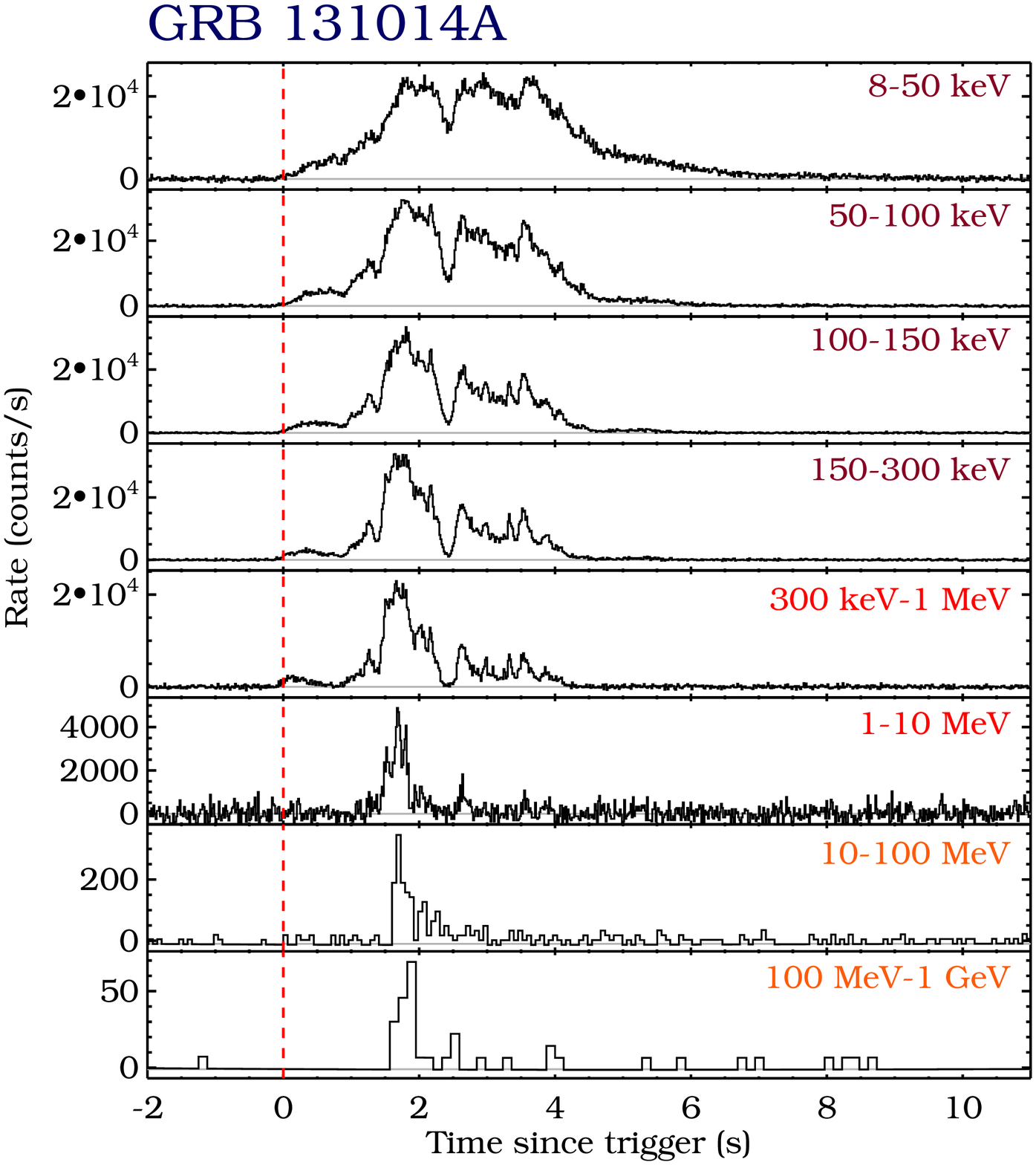} 
\end{tabular}
\caption{Light curves over several energy bands of two GRBs from our
subsample:
GRB 100116A ({\it left panel}) and GRB 131014A ({\it right panel})} 
\label{Fig4}
\end{figure*}

In Figure \ref{Fig4} we plot these significances 
calculated in the four BGO energy bands as a function 
of the LAT boresight angle $\theta$. 
The dashed vertical line indicates the LAT 
FoV at $\theta = 65^{\rm o}$. 
GBM-only detections are marked as gray circles.
Filled and empty circles represent
long and short GRBs.
69 bursts from our sample are also detected by the 
LAT \cite{PUBtab}: 
58 GRBs are detected with the standard likelihood analysis 
above 100 MeV ({\it stars}). Out of those, 33 are
detected also below 100 MeV with the 
LLE technique ({\it squares}). Moreover, there are
11 bursts which are detected only with LLE analysis
({\it triangles}). Long-duration GRBs are
plotted with {\it blue} symbols and short GRBs 
are plotted in {\it red}.

In these proceedings we want to focus only 
on the brightest events of our sample
according to two criteria: (a) the BGO detection 
significance in the 500--1000 keV energy band, 
combined with the significance in the 5--10 MeV 
energy;
and (b) the LLE detection significance in the 10 MeV -- 1 GeV
energy range. The latter is calculated
by means of an algorithm presented in 
Section 3.3.1 of \cite{ACK13} and specifically designed
for LLE source detection. The LLE data presented in this analysis 
are produced from {\it Pass 8} data.

If we independently select the 20 brightest events with 
both criteria, we end up with a subsample of 27 GRBs,
which are listed in Table \ref{Tab1}. There are 21 long
and 6 short GRBs in this new subsample. In the table
we specify the GRB name (column 1), the GBM trigger number and
trigger time in Mission Elapsed Time (MET, columns 2 and 3),
the angle to the LAT boresight $\theta$ (column 4), 
the GBM duration ($T_{\rm 90}$) calculated in the 
50--300 keV energy band and reported by \cite{VON14} (column 5),
the BGO and NaI detectors used for the temporal analysis
(columns 6 and 7), and the detection significances in two
BGO (columns 8 and 9) and in the LLE energy range (column 10).

\subsection{Energy dispersion analysis}\label{SecEnDisp}
In order to select the best energy bands for the temporal
analysis, we first want to study the effect of energy dispersion ($\Delta$E) 
in NaI and BGO data by means of simulations. 
We randomly choose 15 GRBs from 
the latest GBM spectral catalog \cite{GRU13}
and use the best model to simulate their spectra
with XSPEC\cite{XSPEC}. 
Finally, we compare the model predicted rates with the 
measured and simulated rates overs several 
NaI and BGO energy ranges. 

\begin{table}[h]
\begin{center}
\caption{Energy bands of each detector selected for the Duration--Energy analysis}
\begin{tabular}{|c|c|}
\hline \textbf{Detector} & \textbf{Energy bands}
\\
\hline 
\multirow{4}{20mm}{\bf $\quad\;\;\;$NaI} & 8 --50 keV \\
\cline{2-2} 
 & 50--100 keV \\
\cline{2-2} 
 & 100--150 keV \\
\cline{2-2} 
 & 150--300 keV \\
\hline 
\multirow{2}{20mm}{\bf $\quad\;\;\,$BGO} & 0.3 -- 1 MeV \\
\cline{2-2}
 & 1 -- 10 MeV \\
\hline
\multirow{2}{20mm}{\bf $\quad\;\;\;$LLE} & 10--100 MeV \\
\cline{2-2}
 & 100--1000 MeV \\
\hline
\end{tabular}
\label{Tab2}
\end{center}
\end{table}

We find that BGO data show 
an excess count rate in most energy bands ($\sim$30\%), 
worsening towards high energies ($\sim$60\%). 
NaI data show a smaller excess in count rates with 
respect to what is seen in BGOs ($\sim$20\%), 
but in narrow energy bands below 40 keV we see that 
$\Delta$E $\sim$ 30\%. 
In order to keep $\Delta$E $<10$\%\ in NaI detectors and 
$\Delta$E $<20$\%\ in BGO detectors, we decide to selected
the energy bands shown in the Table \ref{Tab2}.
There are four energy bands covered by NaI detectors, two 
covered by BGO detectors and two covered by the LLE technique,
for a total of eight valid spectral bands for the Duration--Energy
relationship analysis.

\begin{table*}[ht!]
\begin{center}
\caption{Sample of 27 bright BGO and LLE GRBs}
\begin{tabular}{|l|c|c|c|c|c|c|c|c|c|}
\hline 
\textbf{GRB }           & 
\textbf{GBM }     & 
\textbf{GBM Trigger}   & 
\textbf{$\theta$ } & 
\textbf{GBM T$_{\rm 90}^a$}   & 
\textbf{BGO} & 
\textbf{NaI } & 
\textbf{$\sigma_{\rm BGO}$} & 
\textbf{$\sigma_{\rm BGO}$} & 
\textbf{$\sigma_{\rm LLE}$} \\
\textbf{Name}   & 
\textbf{Trigger \#}     & 
\textbf{Time (MET)}   & 
\textbf{(Deg.)} & 
\textbf{(s)}   & 
\textbf{det.} & 
\textbf{det.} & 
\textbf{\footnotesize{0.5--1 Mev}} & 
\textbf{\footnotesize{5--10 MeV}} & 
\textbf{\footnotesize{$<$100 MeV}} \\
\hline
080916C	&	080916009	&	243216766.614	&	52.0	&	63.0	$\pm$	0.8	&	0	&	3+4+6	&	7.4	&	6.5	&	65.7	\\
090227B	&	090227772	&	257452263.407	&	72.0	&	1.3	$\pm$	1.0	&	0	&	2+1+0	&	106.6	&	11.7	&	39.9	\\
090228	&	090228204	&	257489602.911	&	16.0	&	0.45	$\pm$	0.14	&	0	&	0+3+1	&	62.0	&	0.0	&	0.0	\\
090510	&	090510016	&	263607781.971	&	13.0	&	0.96	$\pm$	0.14	&	1	&	6+7+9	&	28.5	&	21.5	&	105.0	\\
090902B	&	090902462	&	273582310.313	&	51.0	&	19.33	$\pm$	0.29	&	0+1	&	1+0+9	&	48.0	&	4.0	&	34.3	\\
090926A	&	090926181	&	275631628.987	&	48.0	&	13.76	$\pm$	0.29	&	1+0	&	7+6+3	&	21.1	&	6.8	&	50.3	\\
100116A	&	100116897	&	285370262.242	&	29.0	&	102.5	$\pm$	1.5	&	0	&	0+3+1	&	13.8	&	0.0	&	30.9	\\
100724B	&	100724029	&	301624927.992	&	52.0	&	114.7	$\pm$	3.2	&	0	&	1+0+2	&	14.3	&	5.6	&	82.8	\\
100826A	&	100826957	&	304556304.898	&	71.0	&	85.0	$\pm$	0.7	&	1	&	7+8+6	&	14.2	&	6.3	&	32.1	\\
101014A	&	101014175	&	308722314.622	&	54.0	&	449.4	$\pm$	1.4	&	1	&	7+6+8	&	35.5	&	5.4	&	14.6	\\
101123A	&	101123952	&	312245496.973	&	86.0	&	103.9	$\pm$	0.7	&	1	&	10+9+11	&	15.1	&	7.8	&	29.0	\\
110328B	&	110328520	&	323008161.194	&	31.0	&	141.3	$\pm$	29.8	&	1	&	9+6+0	&	5.6	&	0.0	&	28.9	\\
110529A	&	110529034	&	328322924.872	&	30.0	&	0.51	$\pm$	0.09	&	1	&	9+7+6	&	12.9	&	0.0	&	27.5	\\
110721A	&	110721200	&	332916465.760	&	43.0	&	21.8	$\pm$	0.6	&	1	&	9+6+7	&	30.6	&	18.6	&	135.9	\\
110731A	&	110731465	&	333803371.954	&	6.0	&	7.5	$\pm$	0.6	&	0+1	&	0+3	&	6.9	&	0.0	&	24.4	\\
120817B	&	120817168	&	366868952.723	&	58.8	&	0.11	$\pm$	0.05	&	 1	&	 7+6+8	&	14.2	&	4.0	&	0.0	\\
130305A	&	130305486	&	384176354.369	&	41.4	&	25.6	$\pm$	1.6	&	 1	&	 9+6+0	&	26.4	&	4.0	&	13.0	\\
130310A	&	130310840	&	384638984.503	&	75.9	&	16.0	$\pm$	2.6	&	 1	&	10+9+11	&	45.1	&	43.4	&	26.3	\\
130427A	&	130427324	&	388741629.420	&	47.1	&	138.2	$\pm$	3.2	&	 1	&	 9+10+0	&	472.8	&	23.1	&	40.3	\\
130504B	&	130504314	&	389345526.386	&	61.3	&	0.38	$\pm$	0.18	&	 0	&	 3+4	&	39.0	&	0.0	&	0.0	\\
130504C	&	130504978	&	389402940.518	&	47.5	&	73.2	$\pm$	2.1	&	 1+0	&	 9+0+1	&	25.9	&	6.3	&	31.0	\\
130518A	&	130518580	&	390578080.525	&	40.9	&	48.6	$\pm$	0.9	&	 0+1	&	 3+6+7	&	18.1	&	4.9	&	17.1	\\
131014A	&	131014215	&	403420143.202	&	73.2	&	3.20	$\pm$	0.09	&	1	&	9+10+11	&	148.9	&	11.5	&	31.9	\\
131108A	&	131108862	&	405636118.759	&	24.1	&	18.5	$\pm$	0.4	&	1+0	&	6+3+7	&	13.1	&	0.0	&	61.0	\\
140206B	&	140206275	&	413361375.843	&	46.3	&	116.7	$\pm$	4.2	&	0	&	1+0+3	&	15.7	&	5.0	&	58.4	\\
140306A	&	140306146	&	415769387.951	&	54.7	&	67.3	$\pm$	2.6	&	0	&	3+4+0	&	16.1	&	4.0	&	0.0	\\
140523A	&	140523129	&	422507160.625	&	55.8	&	19.2	$\pm$	0.4	&	0	&	3+4+5	&	26.3	&	0.0	&	4.0	\\
\hline
\multicolumn{10}{l}{\footnotesize{$^{(a)}$ Calculated in the 50--300 keV energy band and reported by \cite{VON14}}.} \\
\end{tabular}
\label{Tab1}
\end{center}
\end{table*}

\section{Temporal analysis}
We select GBM and LLE TTE data
usually binned at 8 ms
in case of short GRBs and enhance the binning up to
64 ms in case of long GRBs.
GBM NaI and BGO detectors are checked for
orientation to the trigger ($<60^{\rm o}$) 
and blockages from the spacecraft.
We then select the three most illuminated NaI detectors 
and one or both BGO detectors (see columns 8 and 9 of
Table \ref{Tab1}).
The GBM energy ranges for the duration analysis
are chosen after the careful check for the detector's
energy dispersion presented in Section \ref{SecEnDisp}. 
The LLE energy range is split into two intervals,
namely 10--100 Mev and 100 MeV--1 GeV.
Errors on the various durations are computed following 
the prescriptions by \cite{KOS96}.
Moreover, systematic errors were computed through an analysis of 
three random samples of bursts, weak, medium, and bright ones:
in each energy band, we changed the various analysis parameters 
(i.e. burst background and light curve
binning selections) and obtained errors of the order of 10--15\%\
in NaI and 20--30\%\ in BGO data. Finally, systematic errors were
add to the statistical ones.

\begin{figure*}[ht!]
\centering
\begin{tabular}{cc}
\includegraphics[width=65mm,bb=0 0 590 482,clip]{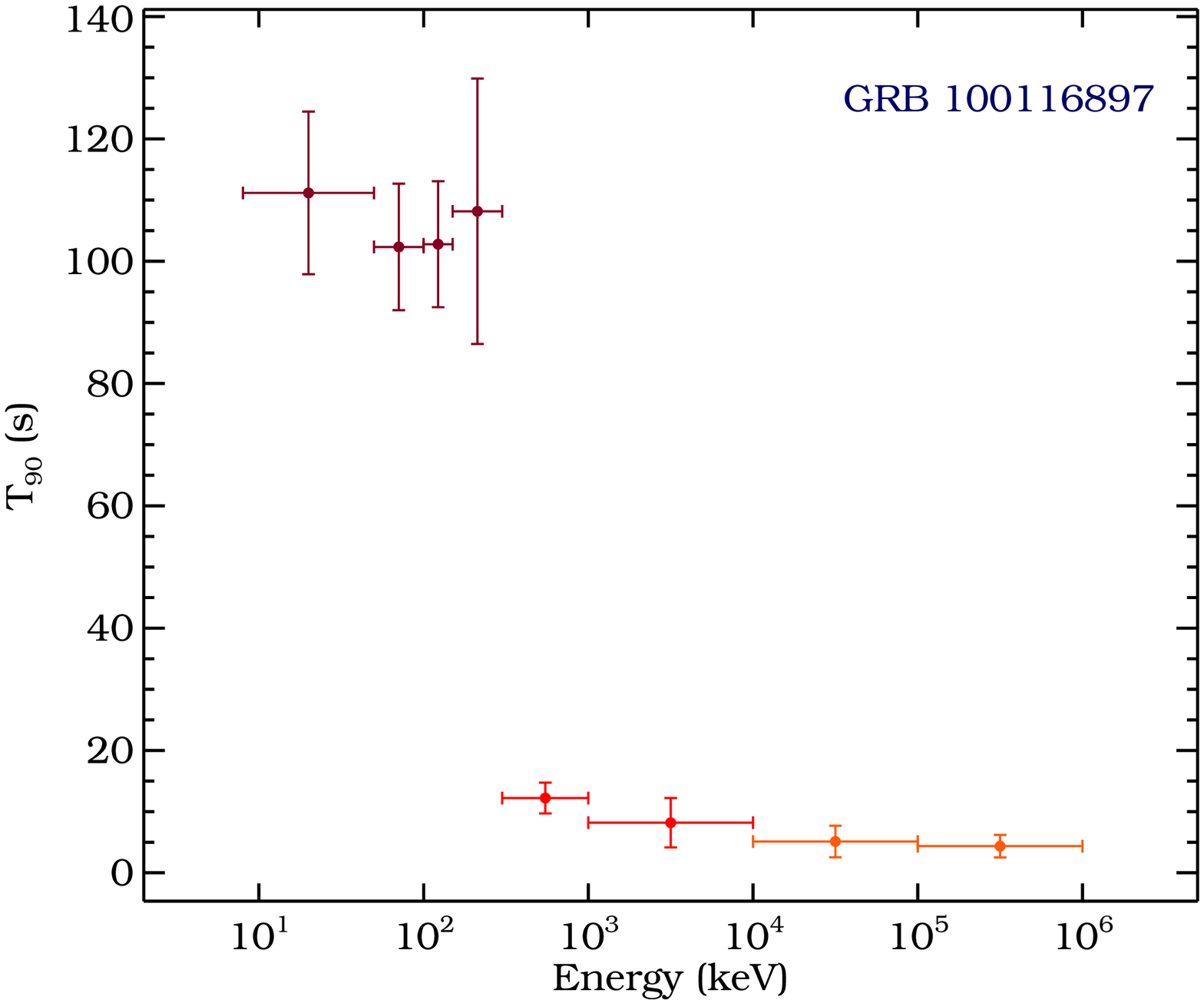} & 
\includegraphics[width=65mm,bb=0 0 590 482,clip]{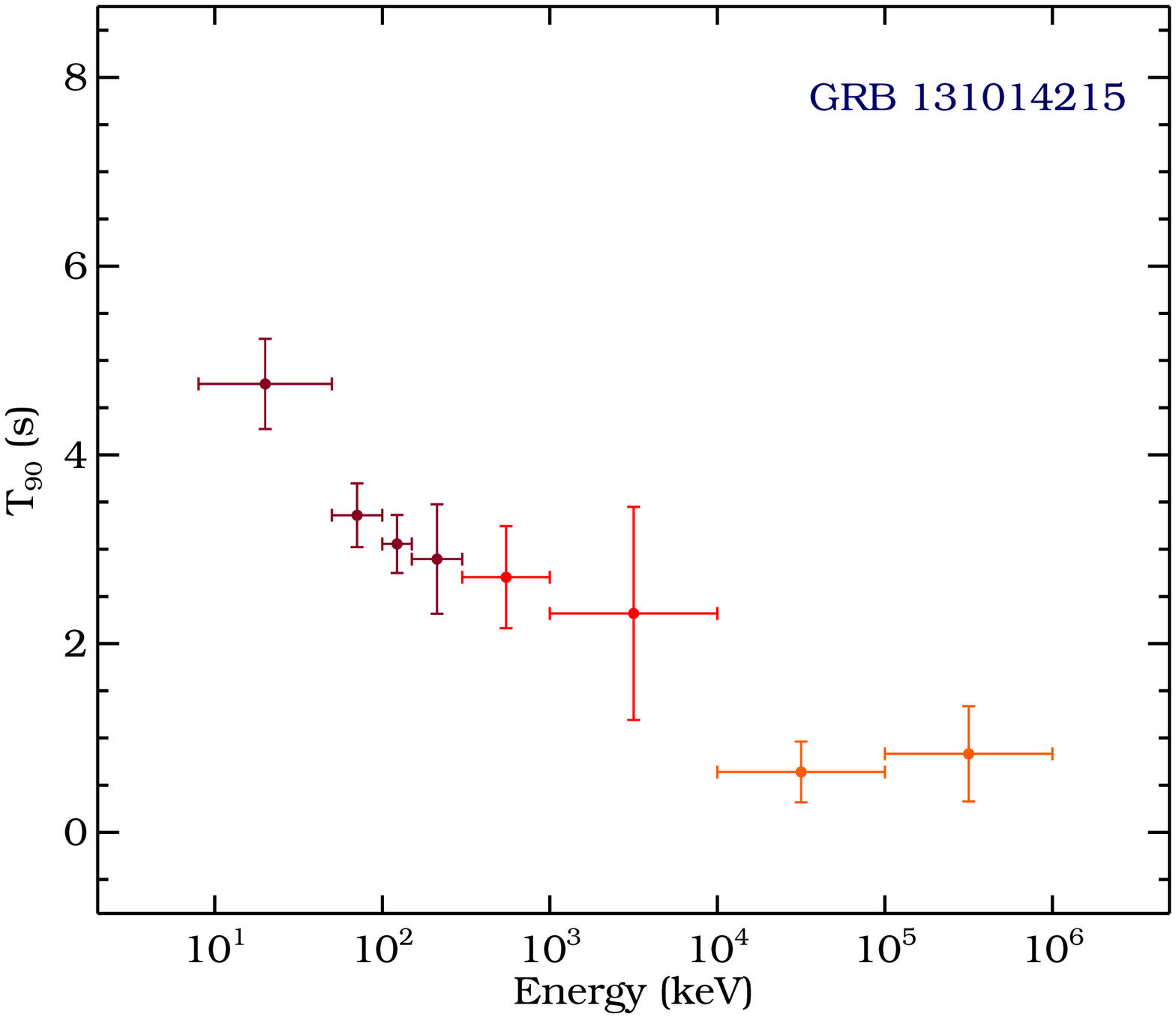} \\
\includegraphics[width=65mm,bb=0 0 590 482,clip]{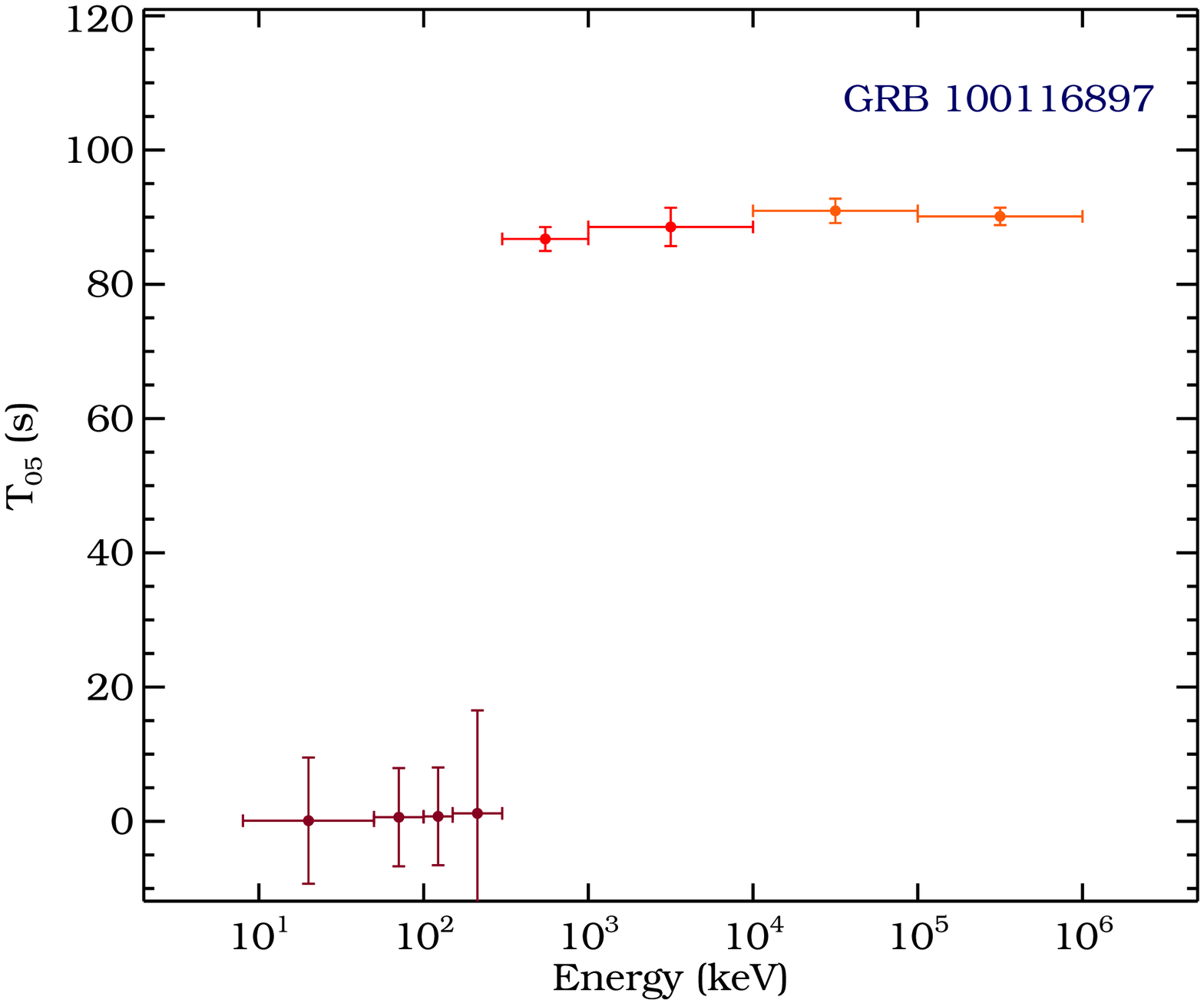} & 
\includegraphics[width=65mm,bb=0 0 590 482,clip]{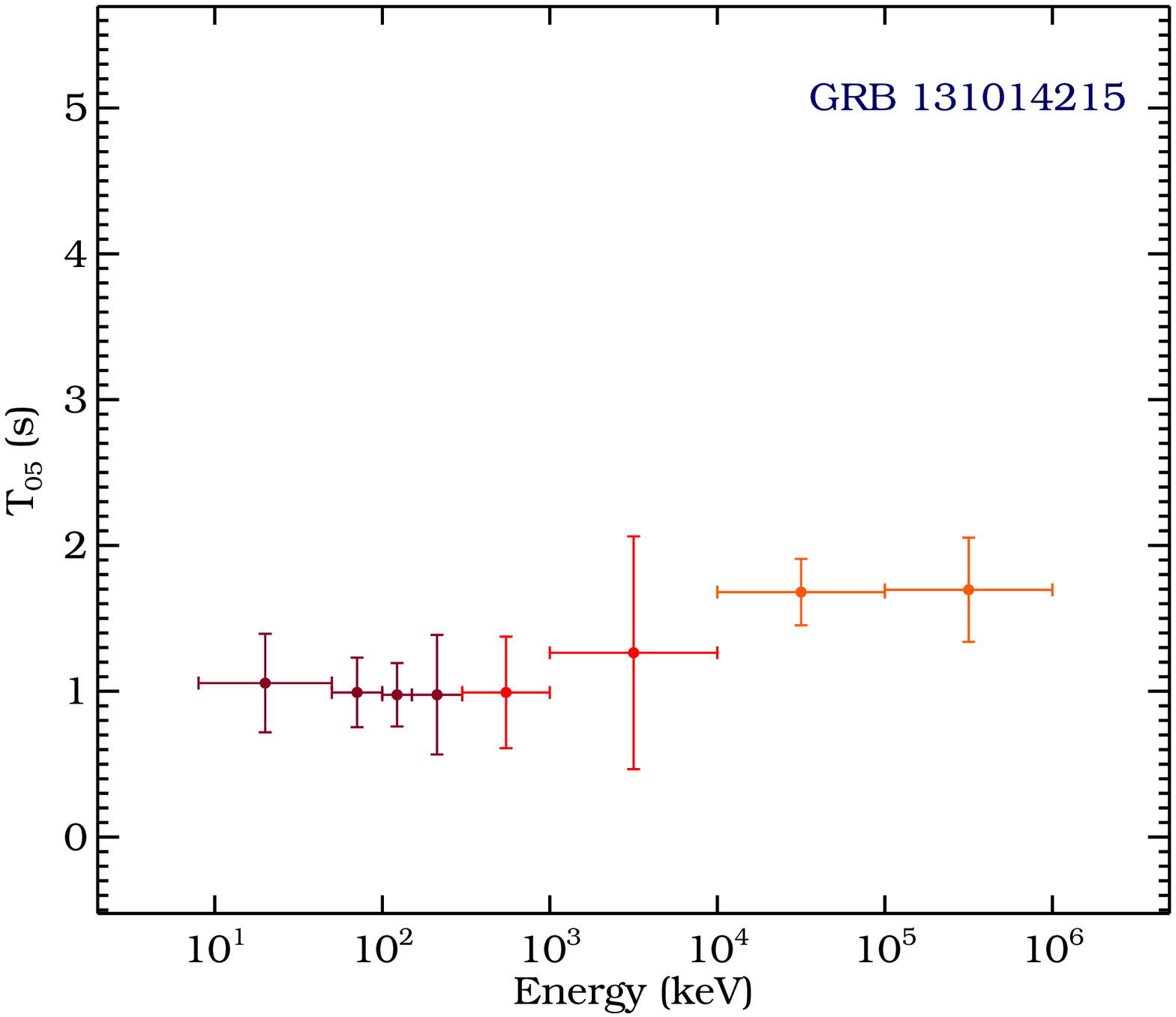} 
\end{tabular}
\caption{Duration--Energy relation ({\it top panels}) and $T_{\rm 05}$--energy relation ({\it bottom panels}) calculated for GRB 100116A
 ({\it left}) and GRB 131010A ({\it right}). Different
colors represent different data used
for the analysis in each energy band as shown in the labels
of Figure \ref{Fig3}.} 
\label{Fig5}
\vspace{10pt}
\includegraphics[width=155mm,bb=0 0 463 164,clip]{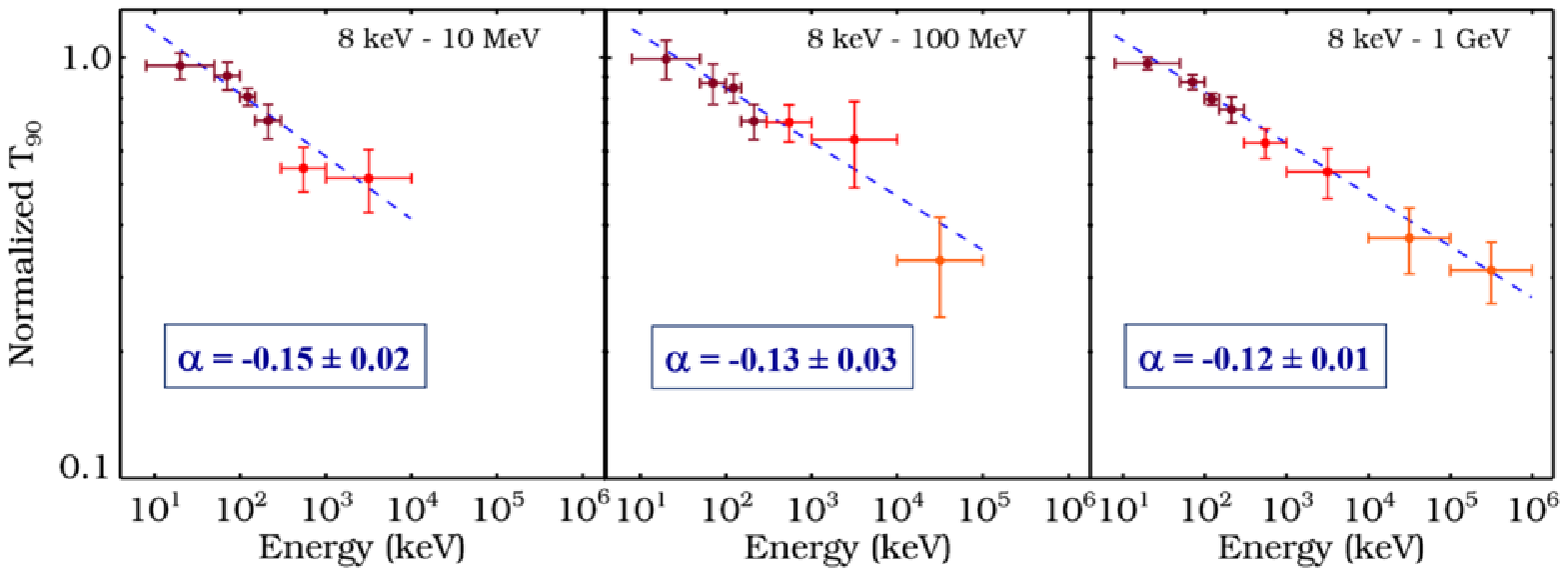} 
\caption{Normalized Duration--Energy relation 
calculated for the 27 bright bursts of our subsample. 
Different colors represent different data used
for the analysis in each energy band as shown in the labels
of Figure \ref{Fig3}.} 
\label{Fig6}
\end{figure*}

Figure \ref{Fig4} shows two example light curves
from GRB 100116A and GRB 131014A.
The trigger time is marked with a vertical
dashed red line and the energy ranges are
labeled in every panel. From the top:
the first four panels represent NaI data ({\it dark red
labels}), the subsequent two panels represent
BGO data ({\it red labels}) while the last panel(s)
represent LLE data ({\it orange labels}).

GRB 100116A  ({\it left panel}) is a rather long GRB, with a two--episode
emission, separated by an 80~s long quiescient
period. The peak at trigger time disappears at 
higher energies. GRB 131014A  ({\it right panel}) 
is much shorter than GRB 100116A and
a delayed start of the high--energy emission 
is evident. This feature is quite common in LAT-detected
bursts (see \cite{ACK13}).

\section{Duration--Energy Relation}
The burst duration  ($T_{\rm 90}$) is calculated by means of 
IDL--based routines and is defined as 90\%\ of the accumulation 
time in count space in each energy band. 
We also calculate $T_{\rm 05}$, which we define as
the beginning of $T_{\rm 90}$ at 5\%\ of counts.
Burst durations and $T_{\rm 05}$ values are
computed in each energy band previously 
defined in Table \ref{Tab2}. 

Results for GRB 100116A and GRB 131010A are plotted in Figure \ref{Fig2}.
The {\it top panels} show the Energy--Duration relation,
while the {\it bottom panels} show the $T_{\rm 05}$--Energy
relation. We adopt different colors for the
data points in order to represent the data from
different detectors ({\it
dark-red:} NaI, {\it medium-red}: BGO; {\it light-red}: LLE)
which were used for the analysis 
in each energy band (as indicated in the labels
of Figure \ref{Fig3}). 
In the case of GRB 100116A, the duration
drops from $T_{\rm 90} \sim 110$ s to just few seconds,
while GRB 131014A's duration smoothly decreases
from one energy band to the next. This effect 
is visible also in the $T_{\rm 05}$ vs.~Energy plots,
where the delayed start of the higher--energy emission
in GRB 131014A is clearly visible.

In order to compare and evaluate the whole subsample of 27 bursts, 
we normalize all $T_{\rm 90}$ measurements and plot 
them as a function of energy in Figure \ref{Fig6}. 
Again, different colors indicate different detectors
used for the temporal analysis.
Since not all bursts in the submsample are seen over all
energy bands, we plot our results in three panels:
7 GRBs in our subsample are detected only up to 10 MeV, so 
no LLE duration could be computed
({\it left panel}). Other 7 GRBs are detected in LLE
but only up to 100 MeV
({\it middle panel}), while 13 GRBs are detected 
all the way up to 1 GeV ({\it right panel}).

We fit the data with a simple power 
law (PL) model ({\it blue dashed lines})
in order to compare the slopes of the 
Duration--Energy relations 
to what is previously reported in the literature.
Results for the PL slope $\alpha$ are reported in
box in the middle of each panel.
\cite{RIC96}, using BATSE data from 25 to $>300$ keV, and 
more recently \cite{BIS11}, 
using GBM BGO data from 300 keV to 10 MeV, 
reported values of the PL slope $\alpha$ between -0.4 and -0.3.
Such values are much steeper then what we find in this analysis. 
Particularly energetic GRBs showing prompt
high--energy emission, i.e. $> 10$ MeV, 
have a much flatter behavior of the Duration--Energy relation.
This possibly indicates that the prompt high--energy
emission is closely related to the low--energy one.
\section{Outlook}
Our future analysis steps include (a)
The comparison of the PL slope $\alpha$ 
of the Duration--Energy relation
deduced from the 27 bright GRBs subsample 
with the one deduced from the 
full sample of 311 bright BGO GRBs;
The correlation of the Duration--Energy relation 
parameters with the burst spectral properties;
and (c) The study of the temporal properties of pulses 
using temporally--resolved spectral analysis for the 
brightest peak of each GRB in the subsample; and
(d) The study of the Duration--Energy relation
at energies $> 1$ GeV, using the LAT standard
analysis. This last step could help 
determining if the highest--energy
emission is in fact
afterglow emission shortly following the start of the
prompt phase emission as seen at smaller frequencies.

%
%

%
%
\end{document}